\documentclass{article}

\usepackage[utf8]{inputenc}     
\usepackage[margin=10pt,font=small,labelfont=bf,labelsep=endash]{caption}
\usepackage[T1]{fontenc}
\usepackage{setspace}
\usepackage{booktabs}
\usepackage{multirow}
\usepackage{tabularx}
\usepackage{paralist}
\usepackage{ifthen}
\usepackage{listings}
\usepackage{amsmath}
\usepackage{amsfonts}
\usepackage{color}
\usepackage[table]{xcolor}
\usepackage{algorithm}
\usepackage{algpseudocode}
\usepackage{url}
\usepackage{rotating}
\usepackage{subfig}
\usepackage{arydshln}
\usepackage{todonotes}
\usepackage[normalem]{ulem}
\usepackage{siunitx}
\usepackage{hyphenat}
\usepackage{authblk}
\usepackage[absolute]{textpos}
\usepackage{xspace}
\usepackage{hyperref} 

\graphicspath{{../thesis/img_out/}{../thesis/img_other/}}
\makeatletter
\def\input@path{{../thesis/}}
\makeatother

\newcommand{\progfunc}{\textnhtt}
\newcommand{\progvar}{\textnhtt}
\newcommand{\swpackage}{\texttt}

\floatstyle{ruled}
\newfloat{proc}{thp}{lop}
\floatname{proc}{Procedure}

\makeatletter
\@ifpackageloaded{algorithm}
	{\algrenewcommand\algorithmicforall{\textbf{for each}}}
	{}
\@ifpackageloaded{amsmath}{
	
	}
	{}
\makeatother

\newcommand{\TikZ}{Ti\textit{k}Z\xspace}

\title{\swpackage{SimOutUtils} - Utilities for analyzing time series simulation output}

\author[1]{Nuno Fachada}
\author[2]{Vitor V. Lopes}
\author[3]{Rui C. Martins}
\author[1]{Agostinho C. Rosa}

\affil[1]{Institute for Systems and Robotics, LARSyS, Instituto Superior Técnico, Universidade de Lisboa, Lisboa, Portugal}
\affil[2]{UTEC - Universidad de Ingenier\'{i}a \& Tecnolog\'{i}a, Lima, Per\'{u}}
\affil[3]{Life and Health Sciences Research Institute, School of Health Sciences, University of Minho, Braga, Portugal}

\providecommand{\keywords}[1]{\textbf{\textit{Keywords---}} #1}

\begin{document}

\begin{textblock*}{210mm}(3mm,3mm)
\noindent The peer-reviewed version of this paper is published in the Journal of Open Research Software (\url{http://doi.org/10.5334/jors.110}). This version is typeset by the authors and differs only in pagination and typographical detail.
\end{textblock*}

\date{}

\maketitle

\begin{abstract}

\swpackage{SimOutUtils} is a suite of MATLAB/Octave functions for studying and analyzing time series-like output from stochastic simulation models. More specifically, \swpackage{SimOutUtils} allows modelers to study and visualize simulation output dynamics, perform distributional analysis of output statistical summaries, as well as compare these summaries in order to assert the statistical equivalence of two or more model implementations. Additionally, the provided functions are able to produce publication quality figures and tables showcasing results from the specified simulation output studies.

\end{abstract}


\section*{(1) Overview}

\subsection*{Introduction}

\swpackage{SimOutUtils} is a suite of MATLAB \cite{matlab2013} functions for studying and analyzing time series-like output from stochastic simulation models, as well as for producing associated publication quality figures and tables. More specifically, the functions bundled with \swpackage{SimOutUtils} allow to:

\begin{enumerate}
\item Study and visualize simulation output dynamics, namely the range of values per iteration and the existence or otherwise of transient and steady-state stages.
\item Perform distributional analysis of focal measures (FMs), i.e. of statistical summaries taken from model outputs (e.g., maximum, minimum, steady-state averages).
\item Determine the alignment of two or more model implementations by statistically comparing FMs. In other words, aid in the process of \textit{docking} simulation models \cite{axtell1996aligning}.
\item From the previous points, produce publication quality \LaTeX\ tables and figures (the latter via the \progfunc{matlab2tikz} script \cite{schlomer2008matlab}).
\end{enumerate}

These utilities were originally developed to study the Predator-Prey for High-Performance Computing (PPHPC) agent-based model \cite{fachada2015template}, namely by statistically analyzing its outputs for a number of different parameters and comparing the dynamical behavior of different implementations \cite{fachada2015template,fachada2015parallelization,fachada2015model}. They were later generalized to be usable with any stochastic simulation model with time series-like outputs. The utilities were carefully coded in order to be compatible with GNU Octave \cite{eaton1997gnu}.

\subsection*{Implementation and architecture}

\begin{figure}[ht]
\centering
\includegraphics[width=\linewidth]{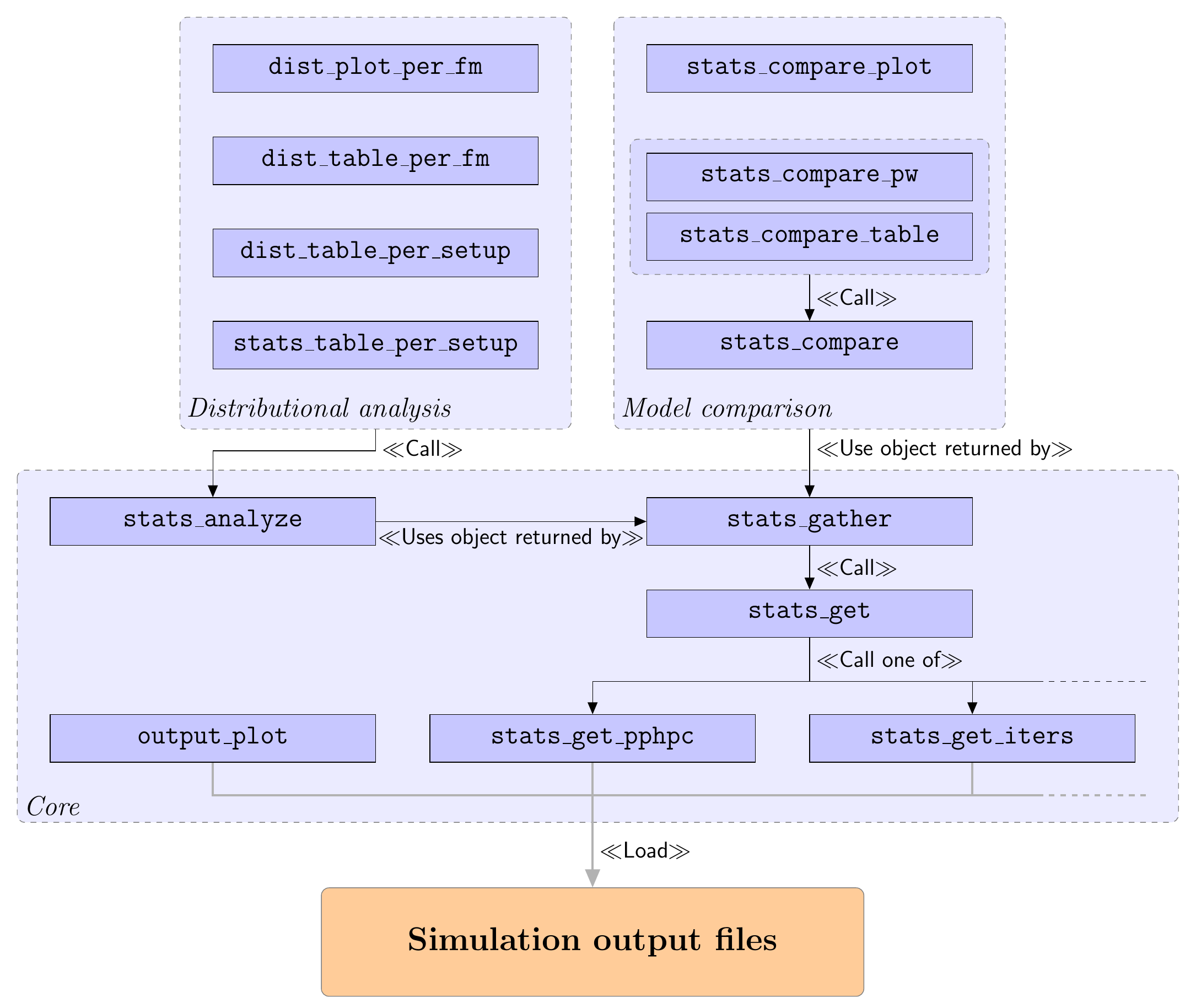}
\caption{\swpackage{SimOutUtils} architecture. Larger blocks with rounded corners and dashed outline constitute function groups, identified in italic font at the lower left corner of the respective block. Within these, functions are represented by smaller blocks with solid outline and sharp corners, with the function name shown in typewriter font. Arrows reflect the relationship between functions and between functions and function groups.}
\label{fig:simoututils_arch}
\end{figure}

The \swpackage{SimOutUtils} suite is implemented in a procedural programming style, and is bundled with a number of functions organized in modules or function groups. As shown in Figure~\ref{fig:simoututils_arch}, the following function groups are provided with \swpackage{SimOutUtils}:

\begin{enumerate}
\item Core functions.
\item Distributional analysis functions.
\item Model comparison functions.
\item Helper and third-party functions (not shown in Figure \ref{fig:simoututils_arch}).
\end{enumerate}

The next sections describe each group of functions in additional detail.

\subsubsection*{Core functions}

Core functions work directly with simulation output files or perform low-level manipulation of outputs. The \progfunc{stats\_get} function is the basic unit of this module, and is at the center of the \swpackage{SimOutUtils} suite. From the perspective of the remaining functions, \progfunc{stats\_get} is responsible for extracting statistical summaries from simulation outputs from one file (i.e., from the outputs of one simulation run). In practice, the actual work is performed by another function, generically designated as \progfunc{stats\_get\_*}, to which \progfunc{stats\_get} serves as a facade for. The exact function to use (and consequently, the concrete statistical summaries to extract) is specified in a namespaced global variable defined in the \swpackage{SimOutUtils} startup script. This allows researchers to extract statistical summaries and use FMs adequate for different types of simulation output.

Two \progfunc{stats\_get\_*} functions are provided, namely \progfunc{stats\_get\_pphpc} and \progfunc{stats\_get\_iters}. The former, set by default, was developed for the PPHPC model, and obtains six statistical summaries from each output: maximum, iteration where maximum occurs, minimum, iteration where minimum occurs, steady-state mean and steady-state standard deviation. It is adequate for time-series outputs with a transient stage and a steady-state stage. The latter, \progfunc{stats\_get\_iters}, obtains statistical summaries corresponding to output values at user-specified instants. It is very generic, and is appropriate for cases where it is hard to derive other meaningful statistics from simulation output. \progfunc{stats\_get\_*} functions are also required to provide the name of the returned statistical summaries. This metadata is used by higher level functions for producing figures and tables.

The \progfunc{stats\_gather} function extracts FMs from multiple simulation output files, i.e., for a number of simulation runs, by calling \progfunc{stats\_get} for individual files. It returns an object containing a $n \times m$ matrix, with $n$ observations (from $n$ files) and $m$ FMs (i.e., statistical summaries from one or more outputs). The returned object also includes metadata, namely a data name tag, output names and statistical summary names (via \progfunc{stats\_get} and the underlying \progfunc{stats\_get\_*} implementation).

The matrix returned by \progfunc{stats\_gather} can be feed into the \progfunc{stats\_analyze} function, which determines, for each sample of $n$ elements of individual FMs, the following statistics: mean, variance, confidence intervals, $p$-value of the Shapiro-Wilk normality test \cite{shapiro1965analysis} and sample skewness. This function is called by all functions in the distributional analysis module, as discussed in the next section.

\begin{figure}[ht]
	\centering
	
	\subfloat[Superimposed, one run.\label{fig:output_plot_a}]{\includegraphics[width=0.48\linewidth]{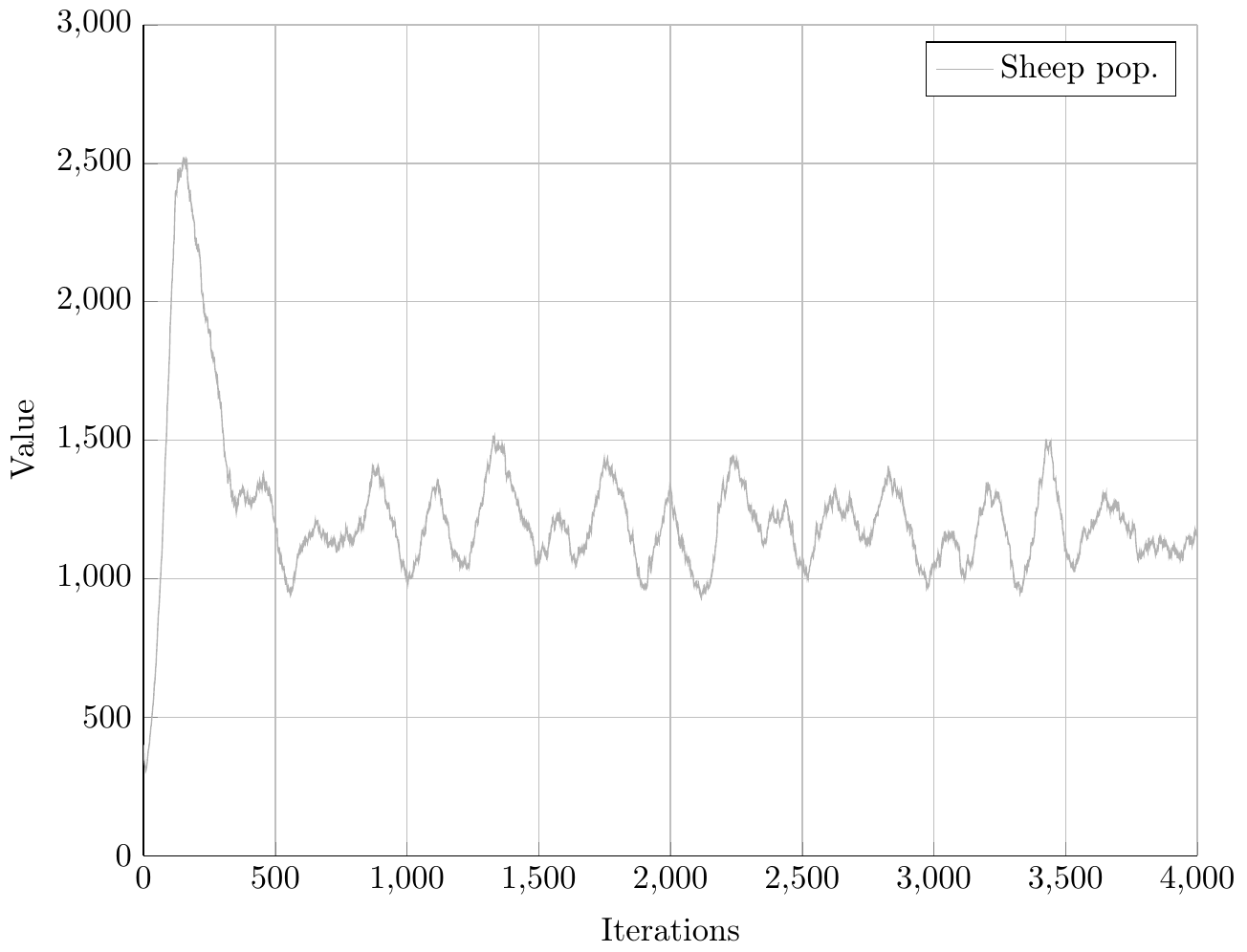}}\quad
	\subfloat[Superimposed, 30 runs.\label{fig:output_plot_b}]{\includegraphics[width=0.48\linewidth]{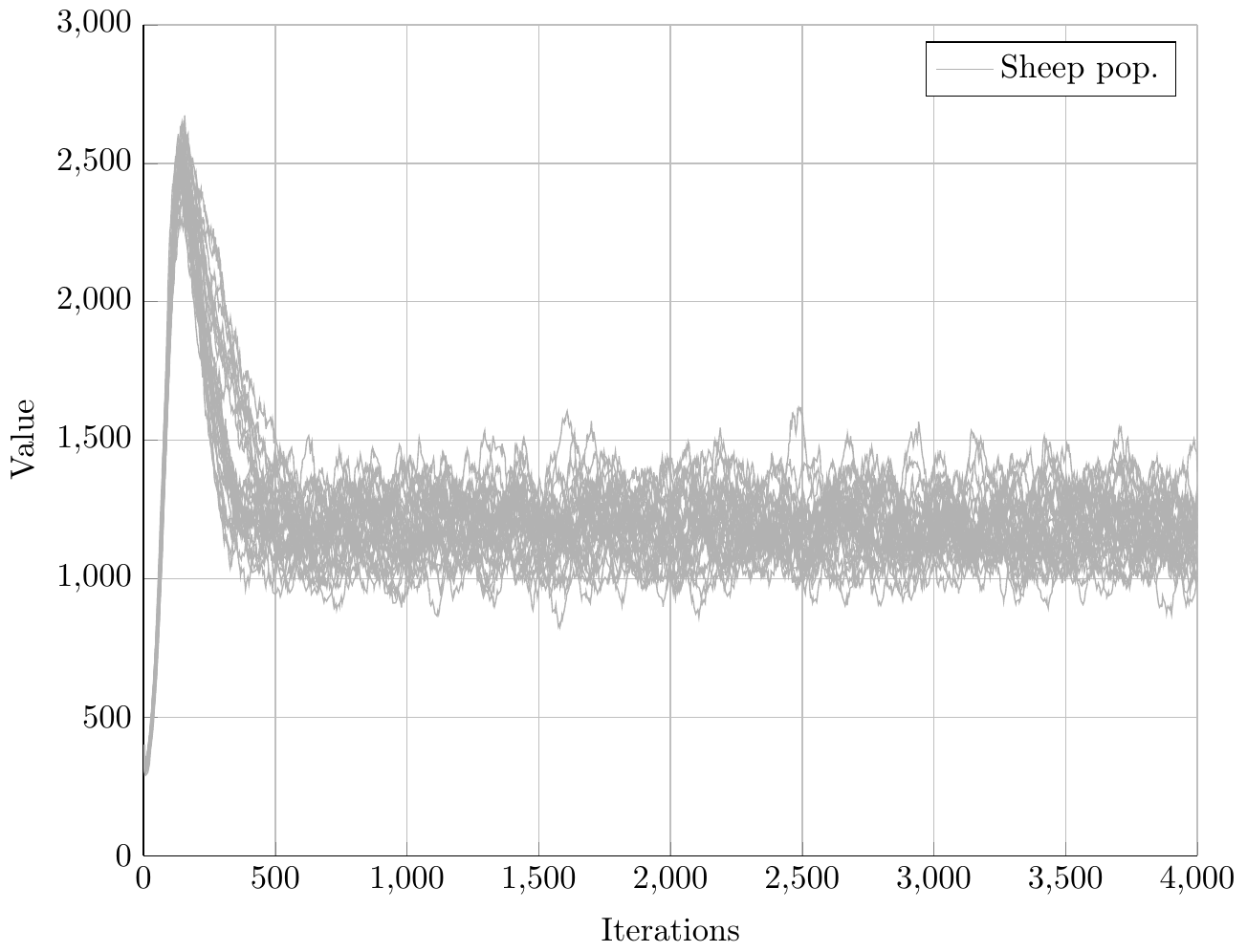}} \\
	\subfloat[Extremes, 30 runs.\label{fig:output_plot_c}]{\includegraphics[width=0.48\linewidth]{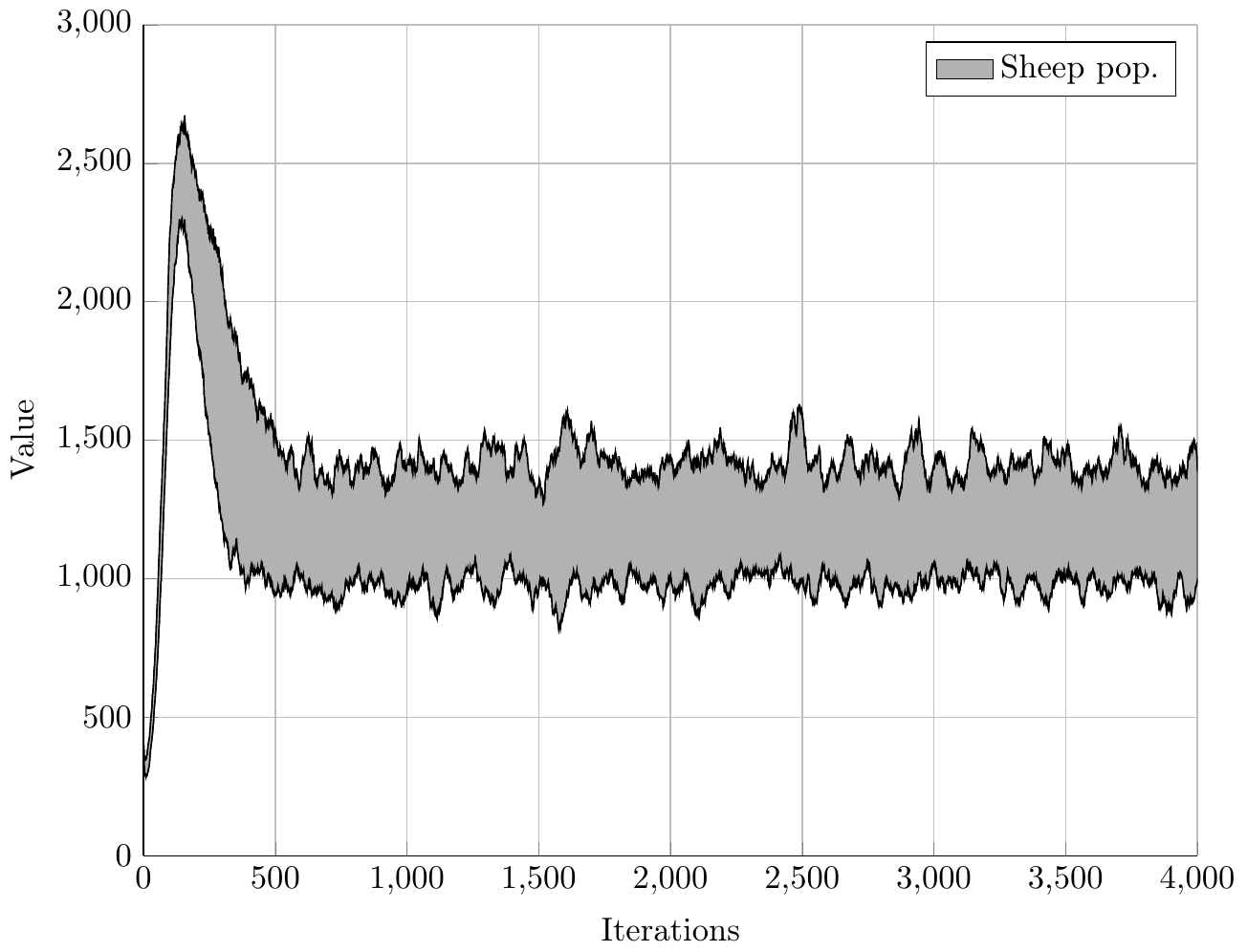}}\quad
	\subfloat[Moving average, 30 runs, window size $=10$.\label{fig:output_plot_d}]{\includegraphics[width=0.48\linewidth]{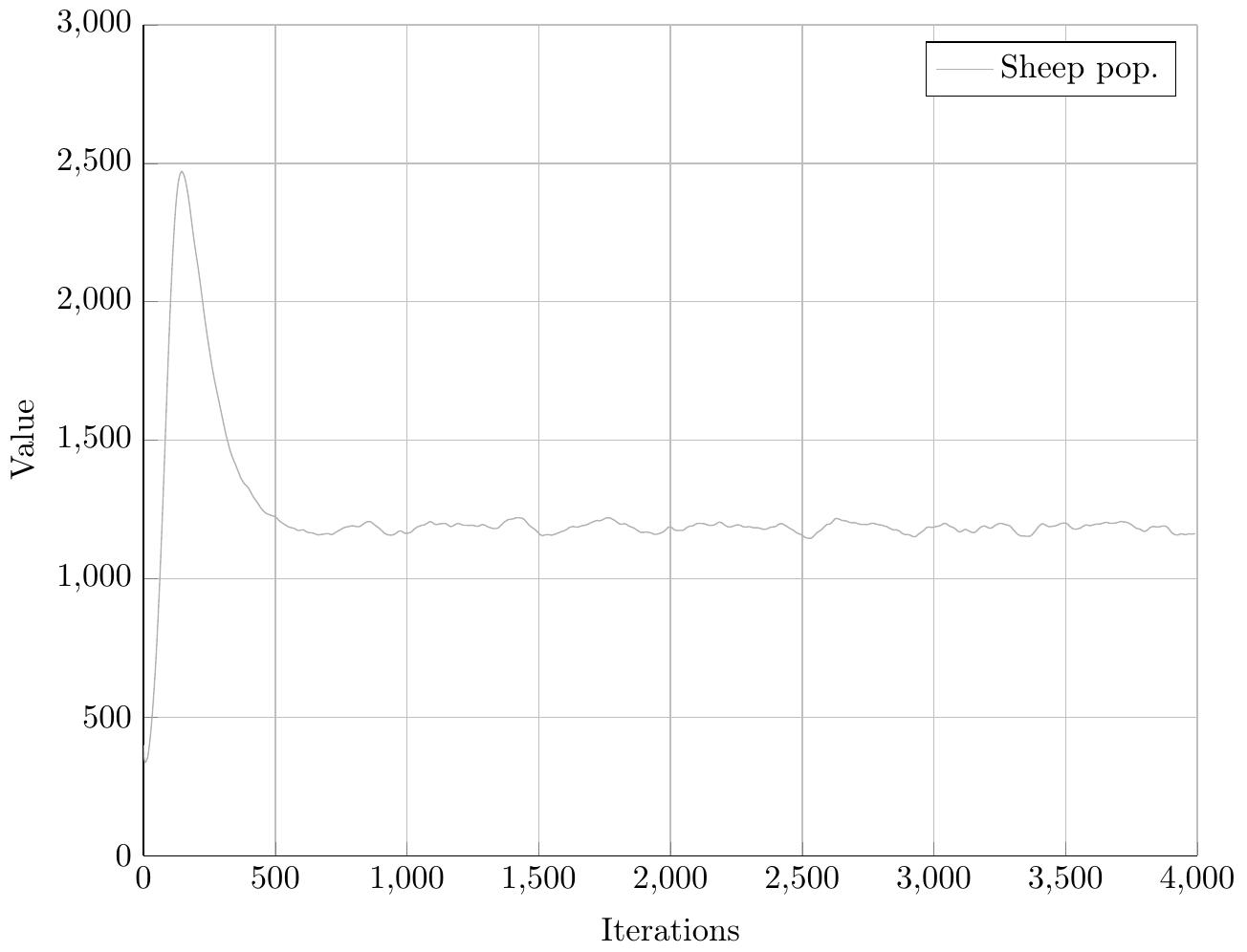}} \\
	\caption{Types of plot provided by the \progfunc{output\_plot} function. All figures show the \textit{sheep population} output from the PPHPC model for size 100, parameter set 1 \cite{fachada2015template}.}
	\label{fig:output_plot}
\end{figure}

Plots of simulation output from one or more replications can be produced using \progfunc{output\_plot}. This function generates three types of plot: superimposed, extremes or moving average, as shown in Figure \ref{fig:output_plot}. Superimposed plots display the output from one or more simulation runs (Figures \ref{fig:output_plot_a} and \ref{fig:output_plot_b}, respectively). Extremes plots display the interval of values an output can take over a number of runs for all iterations (Figure \ref{fig:output_plot_c}). Finally, it is also possible to visualize the moving average of an output over multiple replications (Figure \ref{fig:output_plot_d}). This type of plot requires the user to specify the window size (a non-negative integer) with which to smooth the output. A value of zero is equivalent to no smoothing, i.e., the function will simply plot the averaged outputs. Moving average plots are useful for empirically selecting a steady-state truncation point.

The provided \progfunc{stats\_get\_*} functions, as well as \progfunc{output\_plot}, use the \progfunc{dlmread} MATLAB/Octave function to open files containing simulation output. As such, these functions expect text files with numeric values delimited by a separator (automatically inferred by \progfunc{dlmread}). The files should contain data values in tabular format, with one column per output and one row per iteration.

\subsubsection*{Distributional analysis functions}

Functions in the distributional analysis module generate tables and figures which summarize different aspects of the statistical distributions of FMs. The  \progfunc{dist\_plot\_per\_fm} and \progfunc{dist\_table\_per\_fm} functions focus on one FM and provide a distributional analysis over several setups or configurations, i.e., over a number of model scales and/or parameter sets. On the other hand, \progfunc{stats\_table\_per\_setup} and \progfunc{dist\_table\_per\_setup} offer a distributional analysis of all FMs, fixing on one setup.

The \progfunc{dist\_plot\_per\_fm} function plots the distributional properties of one FM, namely its estimated probability density function (PDF), histogram and quantile-quantile (QQ) plot. The information provided by \progfunc{stats\_analyze} is shown graphically and textually in the PDF plot. The main goal of \progfunc{dist\_plot\_per\_fm} is to provide a general overview of how the distributional dynamics of an FM vary with different model configurations. The \progfunc{dist\_table\_per\_fm} function produces similar content but is oriented towards publication quality materials. It outputs a partial \LaTeX\ table with a distributional analysis for a range of setups (e.g., model scales) and a specific use case (e.g., parameter set). These partial tables can be merged into larger tables, with custom features such as additional rows, headers and/or footers. Tables 8 to 11 of reference \cite{fachada2015template} were generated with this function.

The \progfunc{stats\_table\_per\_setup} function produces a plain text or \LaTeX\ table with the statistics returned by the \progfunc{stats\_analyze} function for all FMs for one model setup. In turn, \progfunc{dist\_table\_per\_setup} generates a \LaTeX\ table with a distributional analysis of all FMs for one model setup. For each FM, the table shows the mean, variance, $p$-value of the Shapiro-Wilk test, sample skewness, histogram and QQ-plot. Supplementary Tables S2.1 to S2.10 of reference \cite{fachada2015template} were created with this function.

\subsubsection*{Model comparison functions}

Utilities in the model comparison group aid the modeler in comparing and aligning simulation models through informative tables and plots, also producing publication quality \LaTeX\ tables containing $p$-values yielded by user-specified statistical comparison tests.

The \progfunc{stats\_compare\_plot} function plots the probability density function (PDF) and cumulative distribution function (CDF) of FMs taken from multiple model implementations. It is useful to visually compare the alignment of these implementations, providing a first indication of the docking process.

The \progfunc{stats\_compare} function is the basic procedure of the model comparison utilities, comparing FMs from two or more model implementations by applying user-specified statistical comparison tests. It is internally called by \progfunc{stats\_compare\_pw} and \progfunc{stats\_compare\_table}, as shown in Figure \ref{fig:simoututils_arch}. The former applies two-sample statistical tests, in pair-wise fashion, to FMs  from multiple model implementations, outputting a plain text table of pair-wise failed tests. It is useful when more than two implementations are being compared, detecting which ones may be misaligned. The latter, \progfunc{stats\_compare\_table}, is a very versatile function which outputs a \LaTeX\ table with $p$-values resulting from statistical tests used to evaluate the alignment of model implementations. It was used to produce Table~8 of reference \cite{fachada2015parallelization} and Table~1 of reference \cite{fachada2015model}.

\subsubsection*{Helper and third-party functions}

There are two additional groups of functions, the first containing helper functions, and the second containing third-party functions.

Helper functions are responsible for tasks such as determining confidence intervals, histogram edges, QQ-plot points, moving averages and whether MATLAB or Octave is being used. Functions for formatting real numbers and $p$-values, as well as for creating very simple histograms and QQ-plots in \TikZ \cite{tantau2013tikz} are also included in this group.

A number of third-party functions, mostly providing plotting features, are also included. The \progfunc{figtitle} function adds a title to a figure with several subplots \cite{greene2013figtitle}. The \progfunc{fill\_between} function \cite{vincent2014fill} is used by \progfunc{output\_plot} for filling the area between output extremes. The \progfunc{homemade\_ecdf} function \cite{boutin2011homemade} is a simple Octave-compatible replacement for the MATLAB-specific \progfunc{ecdf}, assisting \progfunc{stats\_compare\_plot} in producing the empirical CDFs. In turn, the \progfunc{kde} function \cite{botev2010kernel} is used to estimate the PDFs plotted by \progfunc{stats\_compare\_plot} and \progfunc{dist\_plot\_per\_fm}. The \progfunc{swtest} function is the only third-party procedure not related to plotting, providing the $p$-values of the Shapiro-Wilk parametric hypothesis test of normality \cite{saida2007shapiro}. Some of these functions were modified, in accordance with the respective licenses, for better integration with the goals of \swpackage{SimOutUtils}.

\subsection*{Quality control}

All functions have been individually tested for correctness in both MATLAB and Octave, and most are covered by unit tests in order to ensure their correct behavior. The \swpackage{MOxUnit} framework \cite{oosterhof2015} is required for running the unit tests. Additionally, all the examples available in the user manual (bundled with the software) have been tested in both MATLAB and Octave. These examples range from simple usage patterns to the concrete use cases of the articles in which \swpackage{SimOutUtils} was used \cite{fachada2015template,fachada2015parallelization,fachada2015model}.

\subsection*{Issues and support}

Issues or bugs can be filed at \url{https://github.com/fakenmc/simoututils/issues}. Support for \swpackage{SimOutUtils} is provided on best effort basis by emailing the author at \href{mailto:nfachada@laseeb.org}{nfachada@laseeb.org}.

\section*{(2) Availability}

\subsection*{Operating system}
Any system capable of running MATLAB R2013a or GNU Octave 3.8.1, or higher.

\subsection*{Programming language}
MATLAB R2013a or GNU Octave 3.8.1, or higher.

\subsection*{Dependencies}
MATLAB requires the Statistics Toolbox.

\subsection*{List of contributors}
The software was created by Nuno Fachada.

\subsection*{Software location}

\subsubsection*{Archive}

\begin{description}

\item[Name] \swpackage{SimOutUtils}
\item[Persistent identifier] \url{http://dx.doi.org/10.5281/zenodo.50525}
\item[Licence] MIT License
\item[Publisher] Zenodo
\item[Date published] 26/04/2016

\end{description}

\subsubsection*{Code repository}

\begin{description}

\item[Name] \swpackage{SimOutUtils}
\item[Identifier] \url{https://github.com/fakenmc/simoututils}
\item[Licence] MIT License
\item[Date published] 26/04/2016

\end{description}

\subsection*{Language}
English.

\section*{(3) Reuse potential}

These utilities can be used for analyzing any stochastic simulation model with time series-like outputs. As described in `Core functions`, output-specific FMs can be defined by implementing a custom \progfunc{stats\_get\_*} function and setting its handle in the \progvar{simoututils\_stats\_get\_} global variable. The core \progfunc{stats\_gather} and \progfunc{stats\_analyze} functions can be integrated into other higher-level functions to perform operations not available in \swpackage{SimOutUtils}.

\section*{Acknowledgements}

This software uses additional MATLAB/Octave functions written by Chad A. Greene \cite{greene2013figtitle}, Benjamin Vincent \cite{vincent2014fill}, Mathieu Boutin \cite{boutin2011homemade}, Zdravko Botev \cite{botev2010kernel} and Ahmed Ben Sa{\"i}da \cite{saida2007shapiro}.

\section*{Funding statement}

This work was supported by the Fundação para a Ciência e a Tecnologia (FCT) projects UID/\allowbreak EEA/\allowbreak 50009/\allowbreak 2013 and UID/\allowbreak MAT/\allowbreak 04561/\allowbreak 2013, and partially funded with grant SFRH/\allowbreak BD/\allowbreak 48310/\allowbreak 2008, also from FCT. The author Vitor V. Lopes acknowledges the financial support from the Prometeo project of SENESCYT (Ecuador).

\bibliographystyle{nar}

\end{document}